\documentclass{nature}

\title{Free-Space Graphics with Electrically Driven Levitated Light Scatterers.}

\author{Johann Berthelot$^{1}$ \& N. Bonod$^{1}$}
\usepackage{graphicx}
\makeatletter
\let\saved@includegraphics\includegraphics
\AtBeginDocument{\let\includegraphics\saved@includegraphics}
\renewenvironment*{figure}{\@float{figure}}{\end@float}
\makeatother

\begin{document}

\maketitle

\begin{affiliations}
 \item Aix Marseille Univ, CNRS, Centrale Marseille, Institut Fresnel, Marseille, France
\end{affiliations}

\begin{abstract}
Levitation of optical scatterers provides a new mean to develop free-space volumetric displays. The principle is to illuminate a levitating particle displaced at high velocity in three dimensions (3D) to create images based on persistence of vision (POV). Light scattered by the particle can be observed all around the volumetric display and therefore provides a true 3D image that does not rely on interference effects and remains insensitive to the angle of observation. The challenge is to control with a high accuracy and at high speed the trajectory of the particle in three dimensions. Systems that use light to generate free-space images either in plasma or with a bead are strictly dependent of the scanning method used. Mechanical systems are required to scan the particles in the volume which weakens the time dynamics. Here we use electrically driven planar Paul traps (PPTs) to control the trajectory of electrically charged particles. A single gold particle colloid is manipulated in three dimensions through AC and DC electrical voltages applied to a PPT. Electric voltages can be modulated at high frequencies (150 kHz) and allow for a high speed displacement of particles without moving any other system component. The optical scattering of the particle in levitation yields free-space images that are imaged with conventional optics. The trajectory of the particle is entirely encoded in the electric voltage and driven through stationary planar electrodes. We show in this paper, the proof-of-concept for the generation of 3D free space graphics with a single electrically scanned particle.
\end{abstract}

Optical scatterers in levitation reveal as competing alternative to holography to yield volumetric images. Different systems have been proposed to provide free-space 3D imaging\cite{Sahoo2016,Ochiai2015,Saito2008,Kimura2006,Perlin2006}. They still suffer from different limitations such as color limitation (in plasma display mainly) and scanning velocity\cite{Blundell2017}. A major breakthrough was achieved in 2018 by the use of photophoretic traps\cite{Smalley2018}. Such traps use strong gradients in temperature around a lossy particle illuminated by a tightly focused laser beam. This trapping mechanism allows to trap particles in the range of a few micrometers up to hundreds of micrometers. This technique was shown to yield high quality centimeter-sized images. The optical trap is scanned with galvanometric mirrors that rely on a mechanical displacement of optical components. Here we show that immobilized planar point Paul traps can scan particles in a 3D space with AC and DC voltages only. The particle trajectory can be fully driven electrically without requiring any physical displacement of any component of the system.

Radio-frequency (RF) traps have been widely investigated to trap and control ions\cite{Paul1990} and they are emerging in nano and microsciences as versatile tools to manipulate nano/micro-objects\cite{Gregor2009,Kuhlicke2014,Kuhlicke2015,Nagornykh2015,Delord2017,Oppock2017}. They rely on the use of a time varying electrical field applied on a special arrangement of electrodes forming an electrostatic potential featuring a saddle shape. RF traps operate with electrically charged nano/micro objects. This step is usually performed with the use of an electrospray ionization system (ESI)\cite{Gaskell1997}. Conventional RF traps are bulky systems that offer low optical access, require high voltages and specific machining facilities. In the 2000's, the extension of this trap configuration into planar geometry has offered novel opportunities with levitation of electrically charged objects\cite{Pearson2006,Kim2010,Kim2011,Eltony2013,Hoffrogge2011,Pearson2006a}. Besides weaker electric voltages required to trap objects, planar electrodes printed on the same plane offer an open access to easily detect and excite the levitated object with optical beams. Furthermore, PPTs offer a full electrical manipulation without disturbing the trapping process\cite{Kim2010}. Another advantage is related to the fabrication. Such systems are easily fabricated with conventional optical lithography or laser etching. The high stability of PPTs was unveiled in 2016 by translating and rotating PPTs while keeping particles trapped in levitation in air\cite{Alda2016}. We show in this study that PPTs are highly suited to yield free-space images by manipulating and displacing individual optical scatterers by plugging the trajectory of particles into AC and DC voltages applied to stationary PPTs. 

\begin{figure}[h!]
\begin{center}
  \includegraphics[width=\linewidth]{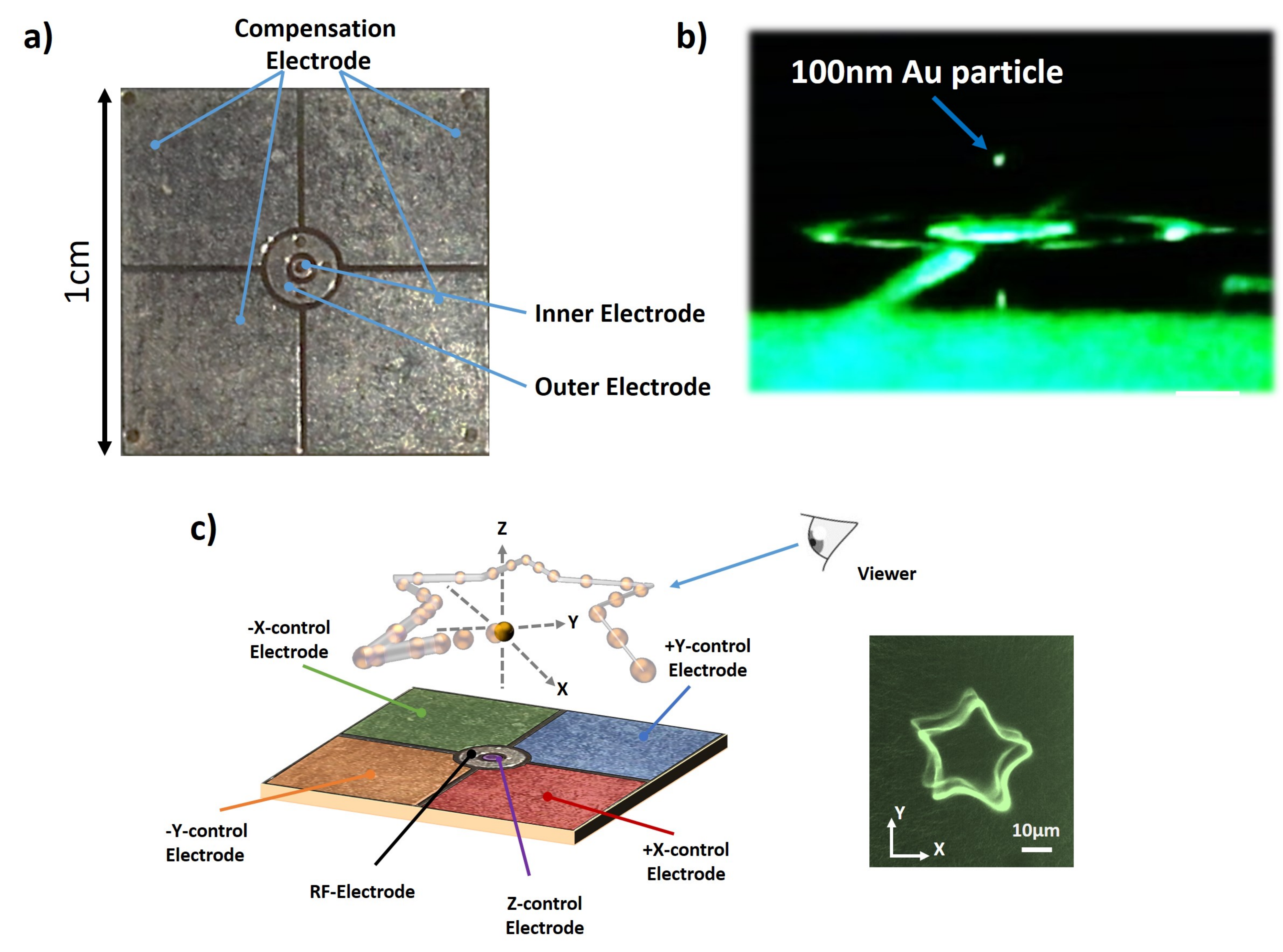}
  \caption{Planar Paul traps. (a) Photography of a surface electrode trap used in this study. It is composed of different electrodes: inner electrode, RF electrode and 4 compensation DC electrodes. (b) Photography of a 100nm gold particle put in levitation with the PPT presented in (a). (c) Schematic illustration of the electrically driven motion of the particles. Each compensation electrode is used to control a $-Ox$, $+Ox$, $-Oy$, $+Oy$ direction. The displacement value is related to the amplitude of the applied electric field. The trajectory along the $z$ axis is controlled with the inner electrode. Insert on the right is a free-space 2D image of a star obtained with our PPT. }
  \label{fgr:fig1}
  \end{center}
\end{figure}

A typical example of PPT is presented in Fig.\ref{fgr:fig1}(a). The dimensions of the different electrodes are 0.7 mm in diameter for the inner electrode and 3 mm in diameter for the outer electrode. They were fabricated on a commercial Printed Circuit Board (PCB). It consists of two inner and outer ring electrodes surrounded by four compensation DC electrodes. The whole PPT has a size of 10 mm. Typically this PPT geometry requires oscillating fields in the kHz range with an amplitude of hundreds of volts. Connections are made underneath the planar PCB. By polarising the different electrodes, either with DC or AC voltage, it is possible to trap a single nano-object with a high stability (cf. Fig.\ref{fgr:fig1}(b)). The four compensation electrodes add external parameters to displace the object in the plane while the inner electrode controls the out-of-plane displacement. The motion of the object is then driven depending of the configuration and the applied voltage amplitude of the different fields, $i.e.$ AC or DC voltage. An important point is that all the AC voltages must be applied with the same frequency\cite{Alda2016}. The circuit configuration used in this study is illustrated in Fig.\ref{fgr:fig1}c. Thanks to this system, the trajectory of the particle can therefore be driven by directly encoding the amplitude in function of the position and applied to the electrodes by analog outputs. The speed and the precision of the driven motion are then strictly dependent of the electronic systems, $i.e.$ bandwidth and resolution. In our experiments we used an electronic card from National Instrument (NI PCIe-7841R).

In the case where no driven electric field is applied, and taking into account the background pressure, the motion of the trapped particle can be predicted by solving the Mathieu equations\cite{Nasse2001}:
\begin{equation}
\frac{d^2u}{dy^2}+\mu \frac{du}{dy}+ (a_u-2q_u\cos2y)u=0,
\label{mathieu}
\end{equation}
where $u$ denotes either the $z$ or the $r$ coordinate of the trapped particle. The dimensionless parameter $y$ is related to time by $y=\Omega t/2$, where $\Omega$ is the frequency of the electrical field. The stable motion of the particle is given by the solutions $r$ and $z$ of the Mathieu equation, Eq.\ref{mathieu}, that depends on the dimensionless parameters $a_u$, $q_u$ and $\mu$:
\begin{eqnarray}
&a_u=16\frac{Q}{M}\frac{V_{DC}}{\Omega^2} f(r,z), q_u=8\frac{Q}{M}\frac{V_{AC}}{\Omega^2} f(r,z), \mu=36\frac{\eta}{\rho d^2\Omega}.
\end{eqnarray}

The function $f(r,z)$ depends on the geometry of the electrodes, $d$ is the diameter of the particle, $\rho$ is the density of the material and $\eta$ the viscosity of the medium. We can remark that both parameters $a_u$ and $b_u$ depend on the charge-to-mass ratio $Q/M$ and that $a_u$ depends on a static voltage applied on the electrode. Particles are commonly trapped with an oscillating field only, meaning that $a_u=0$. $a_u$ and $q_u$ depend on the spatial parameter $u=r,z$ but due to the symmetry of the electrode, $a_z$ and $a_r$ are linked $via$ the relation $a_z=-2a_r$. The same relation occurs for the $q_u$ parameters, $q_z=-2q_r$. Solving Eq.\ref{mathieu} provides the $r$ or $z$ coordinates leading to a stable position of the particle in the trap. For that purpose, a second-order Runge-Kutta method is implemented. This allows us to plot the stability diagram of the PPT as a function of the dimensionless parameters $a_u$, $q_u$ for a given $\mu$. $\mu$ is taken equal to 30 in air and equal to 0 in vacuum. We consider the case of an AC voltage with 1kHz frequency, 70 V amplitude and a 100 nm diameter gold particle (typical value used in the experiment). The solutions $x$ and $z$ are calculated with respect to $a_u$, $q_u$ and the stable solutions are plotted in Fig.\ref{fgr:fig1}(a) as functions of $a$ and $q$.
\begin{figure}[h!]
\begin{centering}
  \includegraphics[width=\linewidth]{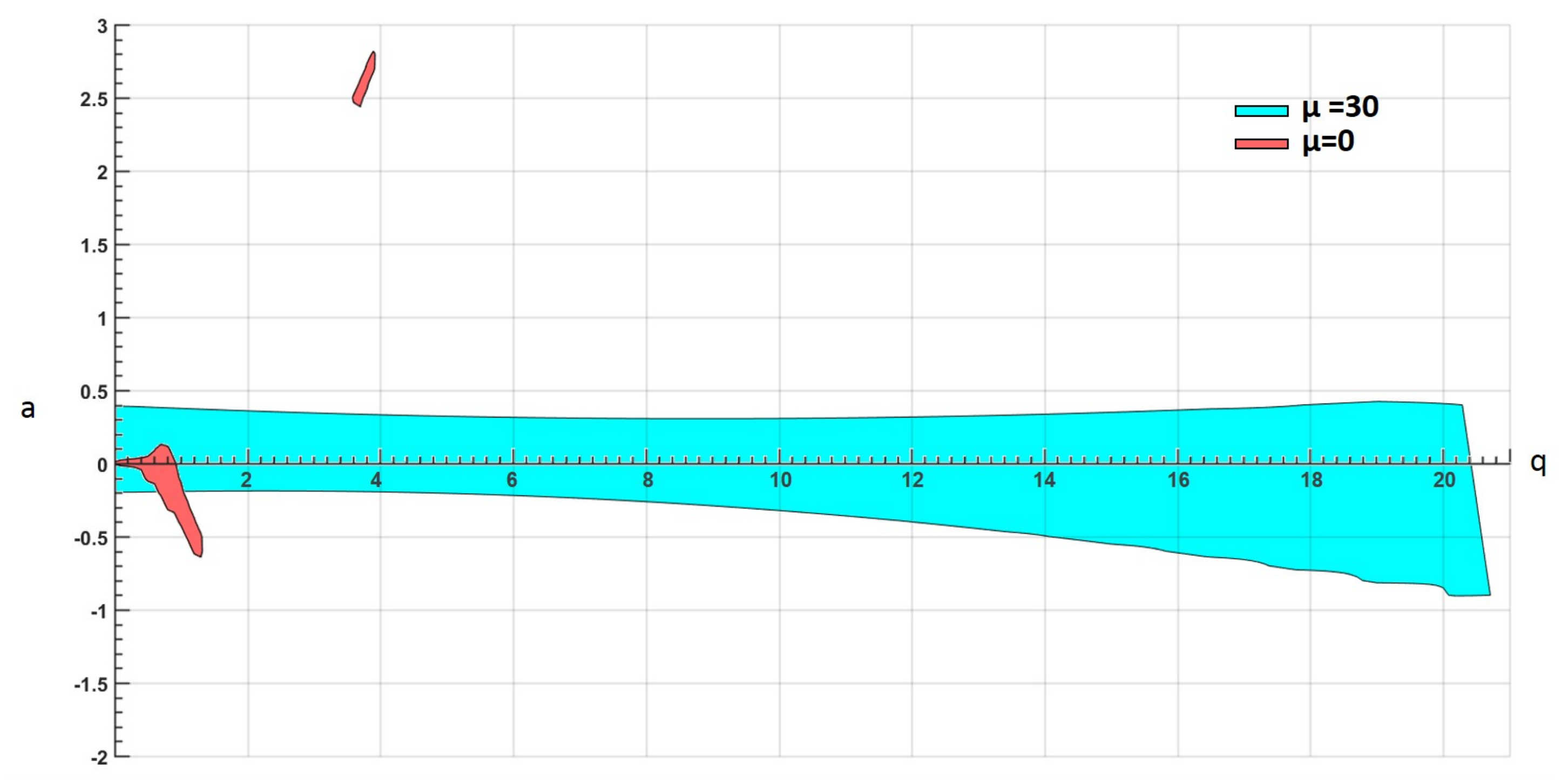}
  \caption{Stability Diagram. Calculated stability diagram for a charged particle in an oscillating electrical field under vacuum (red area) and in air (blue area) conditions. $a$ and $q$ are the solutions of the equation of motion of the particle. Blue areas show stable 3D trapping conditions under air, and red areas under vacuum conditions, both for a 200 nm diameter gold particle.}
  \label{fgr:fig2}
  \end{centering}
\end{figure}
We can clearly see that the domain of stability is much larger in air (in blue) than under vacuum (in red). This result highlights the interest of electrostatic traps to operate in ambient conditions. We can observe in Fig.\ref{fgr:fig2} a large range of stable solutions for a wide range of $a$ and $q$ values, $i.e.$ for a wide range of object sizes with different charge-to-mass ratio $Q/m$, for the same field applied. This also means that RF traps can operate for an even wider range of object sizes by tuning the voltage amplitude and frequency. This numerical study allows to find the voltage parameters that lead to an efficient trap of the gold nanoparticles. It was shown in\cite{Herskind2009} that an additional DC voltage in the radial or axial direction shifts the node of the potential in the radial or axial direction. This property shall be used to shift the position of the node in the radial direction $r$ above the planar electrodes. For that purpose, we use additional DC voltages applied to compensation electrodes. The position of the node in the axial direction can be tuned thanks to the AC voltage. For the radial displacement, the $x$ and $y$ coordinates have first to be discriminated. 4 DC electrodes are placed in the $(0,x,y)$ cartesian coordinates, 2 along the $Ox$ axis at $(-X,0)$ and $(+X,0)$, and 2 along the $Oy$ axis at $(0,-Y)$ and $(0,+Y)$ (see Fig.\ref{fgr:fig3}).
\begin{figure}[h!]
\begin{centering}
  \includegraphics[width=\linewidth]{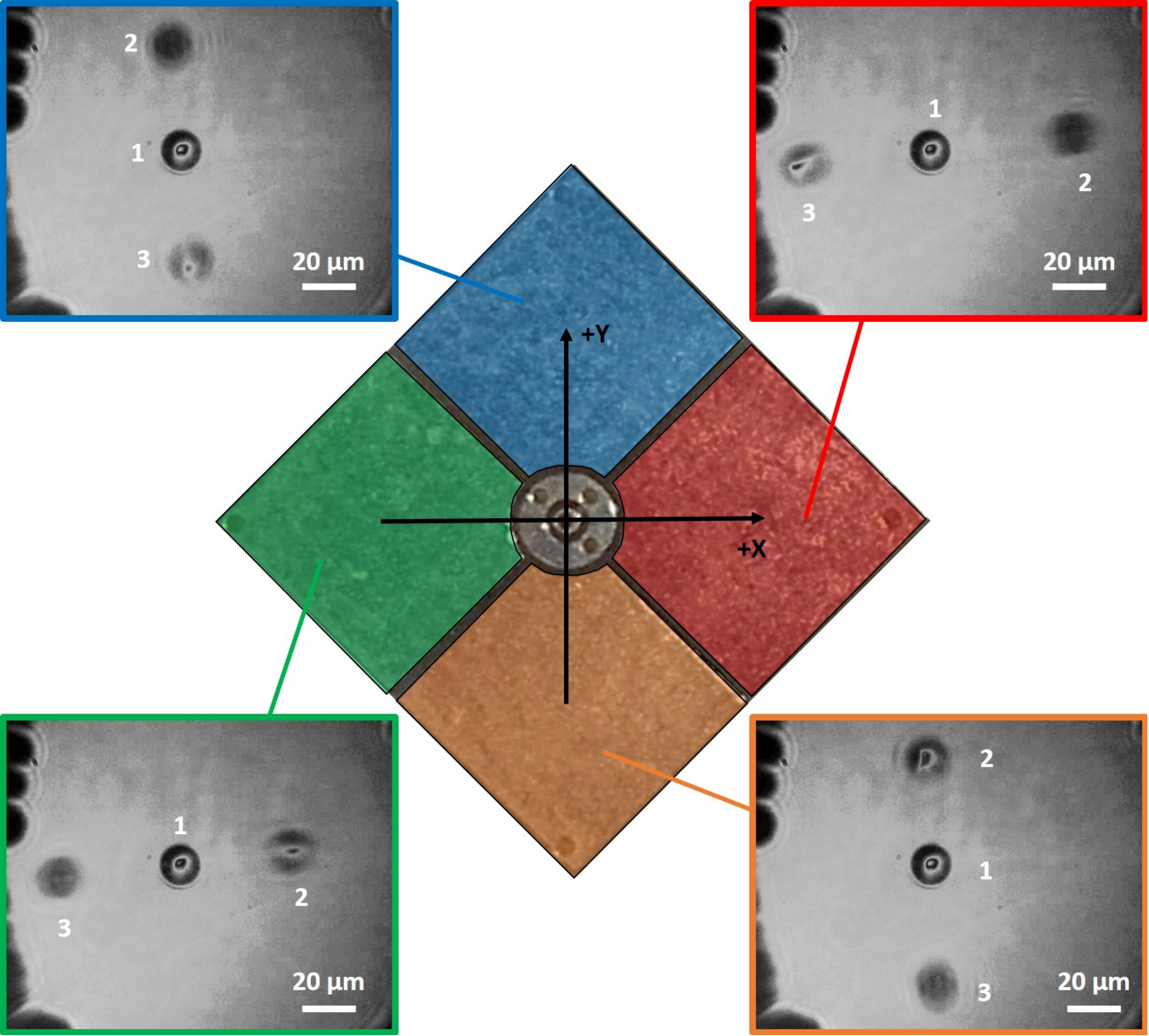}
  \caption{Electrical control of the colloid scatterer motion. Electrical displacement of a 5$\mu$m dielectric particle in function of the voltage applied on the 4 different compensation electrodes. For each electrode, the positions 1, 2, 3 correspond to the voltage: 0V (origin), 10V (+50$\mu$m) and -10V (-50$\mu$m) respectively. }
  \label{fgr:fig3}
  \end{centering}
\end{figure}
A single 5$\mu$m dielectric particle (PMMA) is first trapped by the PPT without any additional DC voltage\cite{Alda2016}. The trajectory of the particle is studied when applying a DC voltage to a single electrode (cf. Fig.\ref{fgr:fig3}(a)). We start with the particle placed at $(0,0)$, $i.e.$ $V_{DC}=0V$ on each compensation electrode. The trajectory of the particle is followed with a CMOS camera (Thorlabs DCC1545M) when applying a DC voltage ranging from -10V to 10V. In our experimental set-up, and in particular with our ESI configuration, the particles are positively charged and they are pushed away linearly from the electrodes. When the DC voltage is decreased back to 0, the particle comes back to its initial position, evidencing the absence of hysteresis. Each DC compensation electrode being driven independently, the particle can be displaced along the $Ox$ or $Oy$-axis. However, in each electrode and when negative DC voltages are applied, we can observe an unfocused effect of the particle resulting from a displacement along the $Oz$ axis. This effect is more pronounced in half direction for each electrode. As presented in Fig.\ref{fgr:fig3}, in the case of the red electrode, the particle is unfocused when a voltage of 10V is applied, ($i.e.$ +$X$ quadrant). It is the inverse effect with the green electrode. This means that the green electrode will be used for displacing the particle in the +$X$-direction and the red one for the -$X$-direction. The same rule will apply for the $Y$-direction. The 4 electrodes define 4 quadrants and therefore allow for a control of the position of the particle in the $(Oxy)$ plane by applying DC voltages only to the compensation electrodes. The displacement of the particle along the $Oz$ axis is controlled $via$ the AC voltage applied to the inner electrode\cite{Alda2016}.

\begin{figure}[h!]
\begin{centering}
  \includegraphics[width=\linewidth]{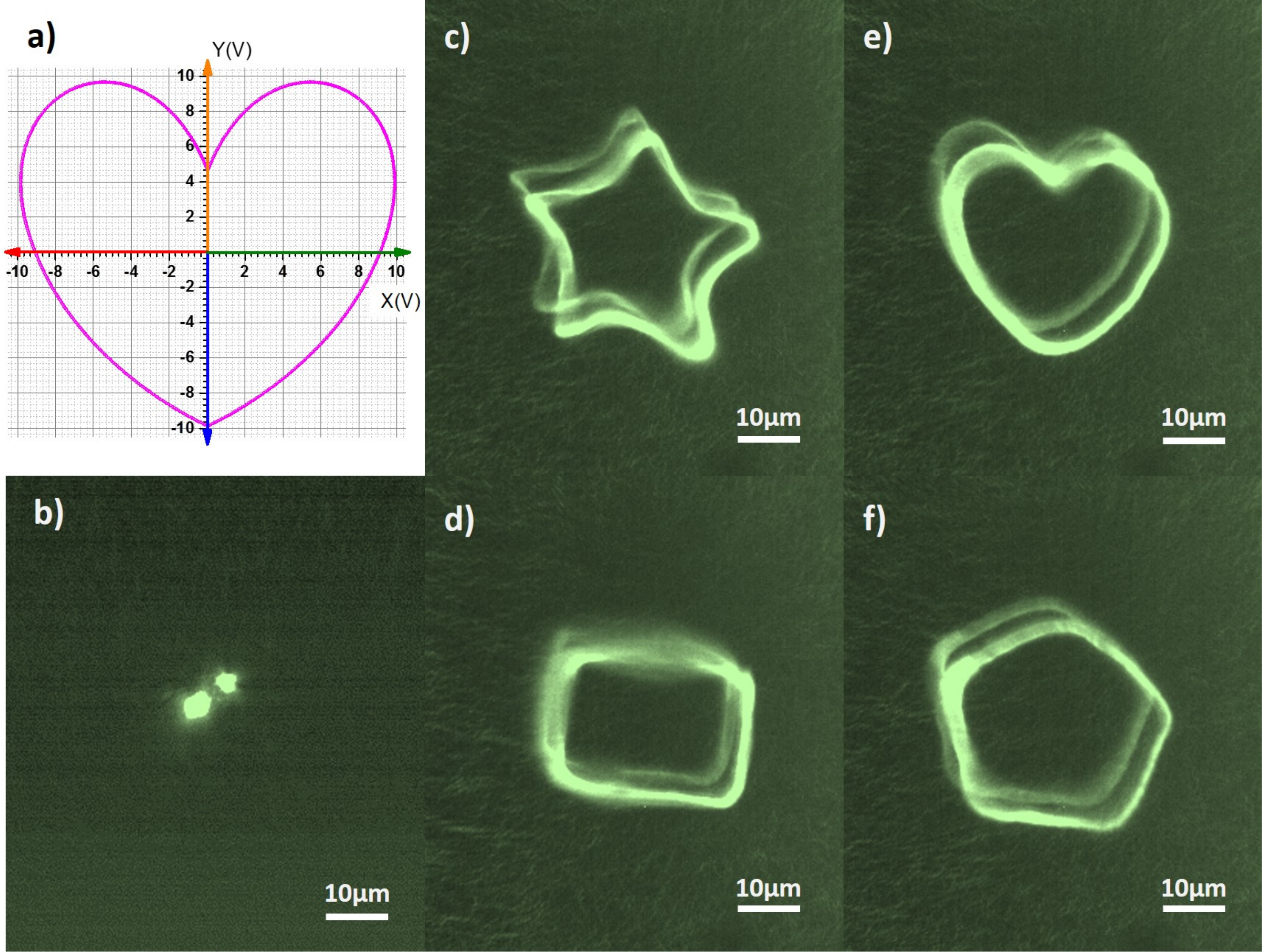}
  \caption{Free space graphics. Mosaic of images showing the performances of the nano-trap display: (a) Plotted trajectory in function of the applied voltage with the identified axes (color corresponding to Fig. \ref{fgr:fig3}). (b) Particle trapped without any motion, c) a star, d) rectangle, e) a heart, f) pentagon. The less intense second trajectory corresponds to an artefact of illumination, linked to the second bright spot in (a). All the geometries have been generated with the same conditions, $V_{AC}=70$V, $f_{trapping}=4$kHz, $V_{DC}=4$V and $f_{sampling}=100$kHz.}
  \label{fgr:fig4}
  \end{centering}
\end{figure}

This electronic system allows for a 3D positioning of the particle with a nanometer precision. Our system provides maximum modulation frequencies up to 180 kHz meaning that the time separating two successive points can be as short as 5.5 $\mu$sec. It exists commercial equipments that provide 1 MHz frequency modulation. We first generate trajectory with a maximum of 4000 points that will behave as pixels of the graphics. The frequency of 100 kHz allows for a scanning of the 4000 pixels of the trajectory in less than 40 milliseconds, $i.e.$ in a time much shorter than the POV (roughly around 100 milliseconds).
Trajectories of the particle in the $(Oxy)$ plane above the planar electrodes are implemented point by point. For that purpose, we used a computing software (Matlab) and the $(x,y)$ coordinates of the contour plot are extracted directly from a black and white images. This list of the $(x,y)$ pixel coordinates are then converted in polar coordinates $(r,\theta)$ and ranked with respect to increasing $\theta$. The ranked coordinates are then converted into voltage values and directly applied to the different electrodes by digital-to-analogue outputs. An example of the treated result with the different directions is presented in Fig.\ref{fgr:fig4}(a). Each $x$ and $y$ coordinate corresponds to a voltage applied to a pair of electrodes of a given quadrant.
This method is followed to generate different images such as stars, hearts, rectangles, hexagons,...(see the video in supplementary). Each trajectory contains 4000 pixels square and is scanned in 40 milliseconds by a single 100 nm gold particle. The particle is illuminated by a collimated green laser beam and the scattered light is collected ($i$) directly by a webcam (logitech C920) and ($ii$) by a microscope objective (Nikon CFI60 TU plan EPI ELWD 50x, NA=0.6) and imaged onto the CMOS camera. The different images generated with the electrically driven levitated plasmonic particle are displayed in Fig.\ref{fgr:fig4} (b-f). The results show the versatility of this electrical approach to tailor the motion of 3D nano-objects at high frequency and accuracy.

In conclusion, we showed the ability of planar RF traps manipulate nanoparticles at high frequency and high precision. The electrode design allows for an independent displacement of the particle in levitation over the 3 cartesian coordinates. The position of the particle can be therefore modified by tuning the applied DC voltage on the four compensation electrodes. The motion dynamics is imposed by the modulation frequency of the electronic card connected to the electrodes (150 kHz). The high frequency allows for driving the particle to 4000 different coordinates in less than 40 milliseconds, a time shorter than the POV. Free-space graphics can therefore be encoded in the trajectory of the particle. The ability of this technique to draw free-space graphics was evidenced by plotting contour plots such as squares, circles, stars,... Our work shows the proof-of-concept of electrical levitation applied to free-space graphics imaging. The volume of the image depends on the typical size of the electrodes and we anticipate that much larger images could be easily achieved ($i$) by increasing the size of the electrodes from the millimeter range to the centimeter size and ($ii$) by using higher voltage amplifiers than the electronic card to the 4 compensation electrodes that will increase significantly the displacement. Planar RF traps are very stable and can trap a wide range of particle sizes, meaning that this technology has a strong potential to yield large size images entirely driven with electrical voltages.










\bibliographystyle{naturemag}


\begin{addendum}
 \item [Acknowledgments]J.B. is supported by Marie Curie fellowship of the European Commission. The authors thank the Fresnel Institute for its financial support through the "Fond pour la Science". The authors thank J\'er\^ome Wenger, Laurent Gallais and Guillaume Baffou for fruitful discussions.
 \item [Authors contribution] J.B. and N.B. formed the original concept and wrote the manuscript. J.B. developed the experimental system, performed all the experiments and the data treatment.
 \item[Competing Interests] The authors declare that they have no competing financial interests.
 \item[Correspondence] Correspondence and requests for materials
should be addressed to J. Berthelot or N. Bonod.~(email: johann.berthelot@fresnel.fr, nicolas.bonod@fresnel.fr).
\end{addendum}

%

\end{document}